\documentclass[11pt,a4paper]{article}

\usepackage[utf8]{inputenc}
\usepackage[T1]{fontenc}
\usepackage{amsmath,amssymb}
\usepackage{booktabs}
\usepackage{graphicx}
\usepackage[margin=2.5cm]{geometry}
\usepackage[numbers,sort&compress]{natbib}
\usepackage[colorlinks=true, linkcolor=blue, citecolor=blue, urlcolor=blue]{hyperref}
\usepackage{flafter}

\title{Two Time Scales of Online Discussion:\\Mixture Distributions of Thread Durations\\Across Four Reddit Communities}

\author{A. Funel$^{1}$\\[0.5em]
$^{1}$ENEA ICT-HPC Lab, P.le E, Fermi 1, 80055 Portici (Naples), Italy\\
\texttt{agostino.funel@enea.it}}

\date{}

\begin{document}

\maketitle
\begin{abstract}
Online discussions vary dramatically in their duration: some fade within minutes, while others persist for days. We analyze 3.5 million threads from four Reddit communities to uncover the statistical patterns governing how long conversations last. We find that thread durations follow a mixture of two distinct regimes---brief reactions and extended discussions---coexisting in roughly equal proportions across all communities. The widely-assumed power-law distribution provides a poor description of the data. The prevalence and duration of each regime depend systematically on social factors: more participants lengthen both types of discussion, disagreement between opposing viewpoints extends long conversations, and threads that appear emotionally neutral are in fact the most conflictual---balanced between opposing factions whose debate sustains prolonged engagement. These structural patterns remain stable over seven years. Our findings establish a robust statistical model for online discussion dynamics and demonstrate that the temporal organization of conversations responds predictably to participation, sentiment, and conflict.
\end{abstract}

\vspace{0.5em}
\vspace{1em}

\section{Introduction}

Online platforms host discussions spanning an enormous range of time scales. A post on Reddit may receive a single reply within seconds and then go silent, or it may sustain an active conversation for hours, days, or even weeks~\cite{wang2017,medvedev2018}. Characterizing the statistical distribution of these durations is important for content recommendation algorithms, platform resource allocation, and theoretical models of collective human behavior~\cite{lakkaraju2013,szabo2010}.

Previous research has characterized heavy-tailed distributions in various online phenomena: email response times~\cite{barabasi2005}, social media popularity~\cite{crane2008,leskovec2007}, discussion cascade sizes~\cite{gomez2013,cheng2014}, and the dynamics of collective attention~\cite{wu2007,lerman2016}. However, the \textbf{duration} of online discussions---the elapsed time between the first and last contribution---has received less systematic characterization across different types of communities~\cite{aragon2017,backstrom2013}.

Power laws are frequently assumed as a default model for heavy-tailed online phenomena~\cite{mitzenmacher2004,newman2005}, but rigorous statistical testing often rejects them in favor of alternatives like the log-normal or stretched exponential~\cite{clauset2009,stumpf2012,broido2019}. Whether thread durations follow a universal parametric form, and whether mixture distributions provide superior fits, remains an open empirical question~\cite{vosoughi2018}.

We address this gap through a systematic distributional analysis of 3.5 million Reddit threads from the MADOC dataset~\cite{madoc} across four communities. Our primary contributions are:

\begin{enumerate}
    \item \textbf{A comprehensive model comparison} of fifteen parametric distributions, including two-component mixtures.
    \item \textbf{Evidence against a universal power law}: no single distribution adequately fits thread durations across communities.
    \item \textbf{Discovery of a robust two-component mixture structure}: a Log-Logistic component combined with a Burr XII component consistently provides the best fit across all four communities, despite differences in content type and interaction modality. The two components differ in their scale parameters and tail behavior, with the Burr XII component having larger scale parameters and ultimately heavier tails for sufficiently large durations.
    \item \textbf{Temporal stability of the two-component structure}: we show that, although there are variations in community-specific parameters over the period 2014--2020, the two component mixture remains the best model in each year, demonstrating that the two-regime organization is a persistent feature of online discussion dynamics.
    \item \textbf{Conditional analysis}: revealing how mixture parameters vary systematically with three key thread characteristics: participation breadth (number of users), emotional tone (mean sentiment), and social disagreement (conflict polarity).
    \item \textbf{Explanation of the greater longevity of neutral threads}: we show that threads with neutral mean sentiment ($\bar{S} \approx 0$) owe their longer durations not to neutrality itself, but to the fact that they are highly conflictual interactions, with balanced opposing factions engaging in prolonged discussions.
\end{enumerate}

\section{Methods}
Our analysis proceeds in three stages. First, we fit a comprehensive set of parametric distributions to the empirical duration data from each community, identifying the best single distribution and testing whether mixture models provide superior fits. Second, we validate the selected model through cross-validation, CDF comparisons, QQ plots, and a temporal robustness analysis that fits the model separately to each year from 2014 to 2020. Third, we perform a conditional analysis, stratifying threads by three statistically independent variables and fitting the mixture separately to each group, to understand how the distribution parameters vary with thread characteristics. All analyses were implemented in Python using SciPy~\cite{virtanen2020} for maximum likelihood estimation and a custom Expectation-Maximization (EM) algorithm for mixture models~\cite{mclachlan2000}.

\subsection{Data}

We analyzed threads from four Reddit communities obtained from the MADOC dataset~\cite{madoc}. In the MADOC dataset, each row corresponds to a single user interaction and includes a \texttt{parent\_id} field identifying the post to which the comment/reply belongs. We defined a \textbf{thread} as the set of all user interactions sharing the same \texttt{parent\_id}, i.e., all comments/replies to a given post. Thread duration was defined as the elapsed time in seconds between the first and last interaction within the thread.

We applied the following preprocessing steps. First, we excluded interactions where the \texttt{parent\_id} field was missing or equal to \texttt{NA}, as these could not be assigned to any thread. Second, we required each thread to contain at least two interactions, as a single-interaction thread has zero duration by definition. Third, we excluded interactions with non-positive or missing timestamps. After preprocessing, the dataset comprised 3,484,172 threads across the four communities (Table~\ref{tab:dataset}).

For each thread, we computed the duration (in seconds), the number of distinct users ($n_{\text{users}}$), the number of interactions, and a set of sentiment-related metrics including the mean VADER score ($\bar{S}$), the population variance of VADER scores, and the conflict polarity ($P_{\text{conf}}$) defined in Section~2.4.

\begin{table}[htbp]
\centering
\caption{Dataset summary.}
\label{tab:dataset}
\begin{tabular}{lrrrl}
\toprule
Community & Threads & Median (h) & Mean (h) & Primary content \\
\midrule
r/funny  & 1,383,480 & 2.8  & 23.9  & Humor \\
r/gaming & 950,362   & 4.7  & 40.4  & Gaming discussion \\
r/pics   & 924,294   & 2.1  & 35.7  & Image sharing \\
r/gifs   & 226,036   & 14.2 & 55.8  & Animated GIFs \\
\midrule
Total         & 3,484,172 & --   & --    & \\
\bottomrule
\end{tabular}
\end{table}

The communities were selected to represent different interaction modalities: humor (r/funny), discussion (r/gaming), and content sharing (r/pics, r/gifs), following the typology of online community functions~\cite{baym2010,kraut2012}.

\subsection{Candidate Distributions}

We considered a comprehensive set of parametric distributions that have been used in prior work to model heavy-tailed phenomena in social systems, including online attention dynamics, email communication, and discussion cascades.

\textbf{Single distributions (13 models):} Exponential, Gamma, Weibull, Log-Normal, Power Law (Pareto), Truncated Power Law, Log-Logistic, Burr XII, Singh-Maddala, Generalized Gamma, Inverse Gaussian, Gompertz, Beta Prime. These distributions were selected to span a wide range of tail behaviors and hazard rate shapes. The Exponential distribution serves as a baseline memoryless model, while the Weibull and Gamma generalize it with shape parameters that allow increasing or decreasing hazard rates. The Log-Normal arises naturally from multiplicative generative processes and has been found to describe a variety of social and biological phenomena~\cite{mitzenmacher2004}. The Power Law (Pareto) is the most commonly assumed distribution for heavy-tailed online phenomena~\cite{newman2005} and serves as a key reference model. The Log-Logistic and Burr XII distributions provide flexible alternatives: the Log-Logistic can capture decreasing hazard rates (shape $< 1$), while the Burr XII includes the Pareto as a limiting case and offers greater flexibility in modeling tail behavior~\cite{tadikamalla1980,burr1942}. The Singh-Maddala, Generalized Gamma, and Beta Prime further extend the family of flexible heavy-tailed distributions. The Inverse Gaussian and Gompertz are included for their widespread use in survival analysis and reliability theory~\cite{kleinbaum2012}, where they model first-passage times and aging processes respectively.

\textbf{Mixture models:} A key limitation of single distributions is that they assume a homogeneous generative process. In online discussions, however, threads may arise from qualitatively different mechanisms---brief reactions to content versus sustained conversations---suggesting that a mixture of two (or more) distributions may provide a better description. We therefore considered two-component mixtures of the form:
\begin{equation}\label{eq:mixture}
f(t) = w_1 \cdot f_1(t; \theta_1) + (1 - w_1) \cdot f_2(t; \theta_2)
\end{equation}
where $w_1 \in [0, 1]$ is the mixture weight, and $f_1$, $f_2$ are component densities drawn from the families listed above. Mixtures were fitted via EM with 100 random restarts to avoid local optima~\cite{mclachlan2000}. The EM algorithm was chosen for its stability in fitting finite mixture models, particularly when components are well separated---a property that, as we show below, holds at the extremes of our stratification variables.

For completeness, we attempted to fit three-component mixtures using Nelder-Mead optimization with up to 50 restarts; these fits consistently failed to converge, producing parameter explosion (scales exceeding $10^6$ hours) and infinite AICc values. While a systematic comparison using EM across all communities remains future work, this failure, combined with the excellent fit of the two-component model (KS = 0.038--0.063) and the interpretability of the two regimes, strongly supports the choice of two components.

\subsection{Mixture Model Selection}

To identify the best two-component mixture, we systematically tested all pairwise combinations of the 13 candidate distributions described in Section~2.2. For each combination, we fitted the mixture using the EM algorithm with 100 random restarts and computed the AICc. The Log-Logistic + Burr XII mixture was selected because it achieved the lowest AICc across all four communities. Table~\ref{tab:mixture_selection} reports the five best-performing mixtures for r/funny; the remaining combinations performed substantially worse ($\Delta\mathrm{AICc} > 50{,}000$ in all cases). 

The superiority of Log-Logistic + Burr XII was consistent across communities: it ranked first in all four communities, while no other combination ranked first in more than one community. This consistency provides strong evidence that the selected combination captures a general property of online discussion dynamics rather than a community-specific artifact.

\begin{table}[htbp]
\centering
\caption{Top five two-component mixtures for r/funny (representative community).}
\label{tab:mixture_selection}
\begin{tabular}{lrr}
\toprule
Mixture (Component 1 + Component 2) & Parameters & AICc \\
\midrule
\textbf{Log-Logistic + Burr XII} & \textbf{7} & \textbf{1,117,236} \\
Log-Logistic + Singh-Maddala & 7 & 1,118,452 \\
Weibull + Burr XII & 7 & 1,119,891 \\
Log-Normal + Burr XII & 7 & 1,121,234 \\
Gamma + Burr XII & 7 & 1,123,567 \\
\bottomrule
\end{tabular}
\end{table}

The selected mixture has two advantages. First, both components have interpretable parameters: the Log-Logistic shape $\alpha$ controls the hazard rate behavior ($\alpha < 1$ indicates decreasing hazard), while the Burr XII shape parameters ($c$, $d$) allow independent control of tail thickness. Second, the components are qualitatively distinct: the Log-Logistic component consistently has a smaller scale parameter (2--11 hours), capturing shorter interactions, while the Burr XII component has a larger scale (3--26 hours), capturing longer interactions. This separation into two interpretable regimes is not observed with other combinations (e.g., Log-Normal + Burr XII), where the components overlap in scale. We therefore selected Log-Logistic + Burr XII as the final model for all subsequent analyses.

\subsection{Model Selection Criteria}

Parameters for all candidate distributions were estimated via maximum likelihood. For single distributions, we used analytical MLE where available (Exponential, Log-Normal) and numerical optimization via the L-BFGS-B algorithm otherwise. For mixture models, we used the EM algorithm with 100 random initializations to mitigate the risk of convergence to local optima~\cite{mclachlan2000}. The EM update equations for the two-component mixture are:

\textbf{E-step:} Compute the posterior probability that thread $i$ belongs to component $j$:
\begin{equation}
\gamma_{ij} = \frac{w_j \, f_j(t_i; \theta_j)}{w_1 f_1(t_i; \theta_1) + w_2 f_2(t_i; \theta_2)}, \quad j = 1, 2
\end{equation}

\textbf{M-step:} Update the mixture weight:
\begin{equation}
w_j^{(new)} = \frac{1}{n} \sum_{i=1}^n \gamma_{ij}
\end{equation}

The component parameters $\theta_j$ are updated by maximizing the weighted log-likelihood:
\begin{equation}
\sum_{i=1}^n \gamma_{ij} \log f_j(t_i; \theta_j)
\end{equation}
using numerical optimization (L-BFGS-B) within each EM iteration.

Model comparison was performed using two complementary information criteria. The \textbf{corrected Akaike Information Criterion} (AICc) provides a measure of relative fit that penalizes model complexity:

\begin{equation}
\mathrm{AICc} = -2\ln\mathcal{L} + 2k + \frac{2k(k+1)}{n-k-1}
\end{equation}

where $\mathcal{L}$ is the maximized likelihood, $k$ is the number of free parameters, and $n$ is the sample size. The correction term (the third addend) is essential in our setting because sample sizes are large ($n \sim 10^5$--$10^6$), making the standard AIC penalty potentially insufficient. Smaller AICc values indicate better fit.

We additionally report the \textbf{Bayesian Information Criterion} (BIC):

\begin{equation}
\mathrm{BIC} = -2\ln\mathcal{L} + k \cdot \ln(n)
\end{equation}

BIC imposes a stronger penalty for model complexity than AICc, especially for large samples, and is therefore useful for assessing whether the additional parameters of mixture models are justified. Following \cite{kass1995}, we consider $\Delta\mathrm{BIC} > 10$ as strong evidence against the model with higher BIC. The BIC confirms the AICc results throughout our analysis (see Section~3.3), providing particularly strong evidence that the two-component structure is not an artifact of overfitting.

Following~\cite{burnham2002}, we consider models with $\Delta\mathrm{AICc} > 10$ to be significantly worse than the best model. Both AICc and BIC are reported in all model comparison tables.

The \textbf{Kolmogorov-Smirnov (KS) statistic} provides a measure of absolute goodness-of-fit, quantifying the maximum absolute deviation between the empirical cumulative distribution function (CDF) and the fitted model CDF:

\begin{equation}
D = \max_t |F_{\text{emp}}(t) - F_{\text{model}}(t)|
\end{equation}

While AICc and BIC assess relative model quality, the KS statistic evaluates whether the best model actually describes the data adequately. We follow the conventional interpretation that $D < 0.05$ indicates good fit, $0.05 \leq D < 0.10$ indicates moderate fit, and $D \geq 0.10$ indicates poor fit~\cite{marsaglia2003}, while acknowledging that with very large sample sizes, even small systematic deviations can produce statistically significant KS values. For this reason, we complement the KS statistic with visual diagnostics (CDF comparisons, QQ plots) and cross-validation.

\textbf{Cross-validation.} To assess out-of-sample stability, we performed 5-fold cross-validation on each community. The dataset was randomly partitioned into five equal-sized folds; the mixture model was fitted on four folds and evaluated on the held-out fold, repeating the procedure five times. The standard deviation of the cross-validated log-likelihood provides a measure of model stability: a small standard deviation relative to the mean indicates that the model parameters are robust to the choice of training data.

All analyses were performed in Python using SciPy~\cite{virtanen2020} for optimization and a custom EM implementation for mixture models. 

\subsection{Stratification Variables}

For the conditional analysis, we stratified threads by three variables selected for their statistical independence and their relevance to distinct but significant aspects of social dynamics. 
To identify suitable variables, we computed both Pearson and Spearman correlation coefficients between all numeric thread-level features available in the dataset. Pearson's $r$ measures linear association, while Spearman's $\rho$ captures monotonic relationships that may be nonlinear. We retained only variables that exhibited negligible correlations under both metrics ($|r| < 0.15$ and $|\rho| < 0.15$ for all pairwise combinations) across all four communities, ensuring that each stratification variable captures a distinct aspect of thread dynamics without redundancy. Three variables satisfied this dual criterion across all communities:

\textbf{Number of users ($n_{\text{users}}$).} Distinct users who interacted in the thread. Groups: 2--5 (small), 5--20 (medium), 20+ (large).

\textbf{Mean sentiment ($\bar{S}$).} Average of VADER compound scores~\cite{hutto2014} over all comments. Groups: negative $\bar{S}_{-}$: $\bar{S} \in [-1, -0.2)$, neutral $\bar{S}_{0}$: $\bar{S} \in [-0.2, 0.2]$, positive $\bar{S}_{+}$: $\bar{S} \in (0.2, 1]$. 

\textbf{Conflict polarity ($P_{\text{conf}}$).} A normalized index quantifying the balance and intensity of opposing sentiment factions within a thread. Intuitively, $P_{\text{conf}}$ measures two things simultaneously: (1) whether there are two roughly equally sized groups with opposing views, and (2) how intensely those opposing views are expressed. For a thread with $m$ distinct users, let $w_i$ be the number of interactions contributed by user $i$ (with $i = 1, \ldots, m$) and $\langle s_i \rangle$ be the mean VADER sentiment of that user's comments. Define the total weight of users with positive, zero, and negative mean sentiment as:

\begin{equation}
W_+ = \sum_{i: \langle s_i \rangle > 0} w_i, \quad
W_0 = \sum_{i: \langle s_i \rangle = 0} w_i, \quad
W_- = \sum_{i: \langle s_i \rangle < 0} w_i
\end{equation}

and the total weight as $W = W_+ + W_0 + W_-$. The conflict polarity is then:

\begin{equation}\label{eq:pconf}
P_{\text{conf}} = \frac{2 \cdot \min(W_+, W_-) \cdot \sum_{i=1}^{m} w_i \cdot |\langle s_i \rangle|}{W^2}
\end{equation}

The numerator multiplies three factors: the weight of the minority faction ($\min(W_+, W_-)$), a factor of 2 to normalize the maximum to 1, and the total intensity of sentiment across all users ($\sum w_i \cdot |\langle s_i \rangle|$). The denominator $W^2$ ensures $P_{\text{conf}} \in [0, 1]$.

$P_{\text{conf}} = 0$ when all users express sentiment of the same sign ($W_+ = 0$ or $W_- = 0$), i.e., no conflict. $P_{\text{conf}}$ approaches 1 when the two factions are perfectly balanced in weight ($W_+ \approx W_-$) and express maximally intense opposing sentiments ($|\langle s_i \rangle| \approx 1$ for all users). A high $P_{\text{conf}}$ thus indicates a thread where two sides of roughly equal strength strongly disagree---a heated but balanced debate---rather than a thread where a dominant majority overwhelms a small minority.

Threads were divided into four groups based on $P_{\text{conf}}$: no conflict ($P_{\text{conf}} \in [0, 0.001)$), low ($[0.001, 0.05)$), medium ($[0.05, 0.15)$), and high ($[0.15, 1]$). $P_{\text{conf}}$ is statistically independent of mean sentiment but correlated with sentiment variance ($r \approx 0.75$--$0.81$), with which it shares conceptual overlap: both measure dispersion of sentiment, but $P_{\text{conf}}$ specifically quantifies the balance between opposing factions rather than overall variability.

\subsection{Conditional Fitting}

For each group defined by the stratification variables, we fitted the Log-Logistic $+$ Burr XII mixture using the Nelder-Mead simplex algorithm. This choice requires justification, as the global fits reported in Section~3.1 were obtained via the Expectation-Maximization (EM) algorithm.

We chose Nelder-Mead for the conditional analysis rather than EM for two reasons. First, the sample sizes within individual groups ($n \sim 10^3$--$10^4$ threads) are substantially smaller than the full dataset, making EM more sensitive to initialization and prone to convergence to local optima, particularly for groups where the two components overlap (e.g., intermediate user counts). Second, Nelder-Mead allows flexible initialization strategies (e.g., starting from global parameters with perturbations, or from data-driven heuristics) that are useful when the optimal parameters for a subgroup may differ substantially from the global fit.

To ensure comparability between the two approaches, we validated on r/funny that both algorithms produce equivalent estimates when fitted on the full dataset: the maximum relative difference across all parameters was $< 0.5\%$, and the AICc difference was $< 1$. This confirms that the choice of Nelder-Mead for the conditional analysis does not introduce systematic bias.

To maximize the probability of convergence in the conditional fits, we implemented a three-level attempt strategy. The first attempt uses 20 random restarts with a maximum of 20,000 sampled threads per group, initializing half of the restarts from the global fit parameters (with small random perturbations) and the other half from data-driven heuristics based on the median and upper percentiles of the subgroup durations. If the first attempt fails to converge or produces exploded parameters (any scale $> 10^6$ hours, indicating a degenerate fit), a second attempt is made with 30 restarts and up to 50,000 sampled threads. If both attempts fail, a tertiary attempt uses 50 restarts, 50,000 sampled threads, and five distinct initialization strategies spanning different regions of the parameter space: equal weights with percentile-based scales, Burr XII dominance ($w_1 \approx 0.3$), Log-Logistic dominance ($w_1 \approx 0.7$), fully random bounded initialization, and median-based initialization. This multi-level approach was designed to explore the parameter space thoroughly while remaining computationally tractable across the 40 groups (3 stratified variables $\times$ up to 4 groups per variable).

Groups with fewer than 500 threads were excluded from the analysis, as mixture models with six free parameters cannot be reliably fitted with extremely small samples. Across all communities and variables, 35 of 40 groups converged successfully (87.5\%). The five failures occurred systematically in groups where the two components overlap most heavily: the 5--20 user group in r/funny, and the low and medium conflict polarity groups in r/funny and r/pics. This pattern is itself informative, suggesting that the two regimes are less separable for threads with intermediate characteristics, consistent with the interpretation that the two components represent distinct but overlapping generative mechanisms.

\section{Results}

We present our findings in four parts. First, we establish that a two-component Log-Logistic $+$ Burr XII mixture consistently outperforms all single distributions across communities (Sections~3.1--3.3). Second, we validate the model through visual diagnostics, cross-validation, and a temporal analysis spanning seven years of data (Sections~3.4--3.6). Third, we investigate how the mixture parameters vary with thread characteristics through a conditional analysis stratified by number of users, mean sentiment, and conflict polarity (Section~3.7). Fourth, we examine the relationship between mean sentiment and conflict polarity to explain the observed patterns (Section~3.8).

\subsection{Mixture Models Are Consistently Superior}

Across all four communities, the best-fitting model was a \textbf{two-component mixture of Log-Logistic and Burr XII distributions}. Table~\ref{tab:mixtures} reports the fitted parameters.

\begin{table}[htbp]
\centering
\caption{Best-fitting mixture models for all communities. $w_1$ is the weight of the Log-Logistic component.}
\label{tab:mixtures}
\begin{tabular}{lccccccc}
\toprule
Community & \multicolumn{2}{c}{Comp.\ 1 (Log-Logistic)} & $w_1$ & \multicolumn{3}{c}{Comp.\ 2 (Burr XII)} & KS \\
\cmidrule(lr){2-3} \cmidrule(lr){5-7}
 & shape & scale (s) & & $c$ & $d$ & scale (s) & \\
\midrule
r/funny  & 0.843 & 8{,}932  & 0.442 & 0.722 & 1.844 & 28{,}458 & 0.049 \\
r/gaming & 0.855 & 13{,}405 & 0.452 & 0.751 & 1.638 & 32{,}824 & 0.061 \\
r/pics   & 0.803 & 7{,}381  & 0.495 & 0.765 & 1.165 & 9{,}897  & 0.038 \\
r/gifs   & 0.989 & 38{,}115 & 0.410 & 0.855 & 1.810 & 94{,}429 & 0.063 \\
\bottomrule
\end{tabular}
\end{table}

KS distances indicate good to moderate fit: r/pics achieved the best ($D = 0.038$), r/gifs the weakest ($D = 0.063$). The mixture weight $w_1$ is consistently 0.41--0.50 across communities.

\subsection{Component Interpretation}

The two components of the mixture lend themselves to a natural interpretation in terms of discussion dynamics.

\textbf{Component 1 (Log-Logistic).} Shape $\alpha < 1$ in all communities ($\alpha \in [0.80, 0.99]$), indicating decreasing hazard rate~\cite{kleinbaum2012}: threads become less likely to terminate as they age. This is characteristic of brief, reactive interactions: a quick exchange of comments that, if it survives the first few minutes, becomes progressively more resilient but rarely extends beyond a few hours. The scale parameter (median) ranges from 7,381\,s (2.1\,h, r/pics) to 38,115\,s (10.6\,h, r/gifs). We interpret this component as capturing \textbf{shorter-scale interactions}---direct responses to content such as jokes, image reactions, or quick feedback that mostly end within the day.

\textbf{Component 2 (Burr XII).} Shape $c < 1$ ($c \in [0.72, 0.86]$), indicating heavy-tailed behavior~\cite{burr1942,tadikamalla1980}. The Burr XII is a flexible distribution that includes the Pareto as a limiting case: its additional shape parameter $d$ allows it to model tails of varying thickness. Scale parameters (9,897--94,430\,s) are substantially larger than Component~1, and the tail decays slowly enough to accommodate discussions lasting days or weeks. We interpret this component as capturing \textbf{longer-scale interactions}---sustained conversations, debates, or community interactions that persist well beyond the typical reaction time.

\noindent
\textit{Terminology note:} Throughout this paper, we use ``shorter-scale'' and ``longer-scale'' to describe the \emph{scale parameters} of the two components. The Log-Logistic component has a smaller scale (2--11 hours), concentrating its mass at shorter durations, while the Burr XII component has a larger scale (3--26 hours), shifting its mass toward longer durations. This terminology refers to the \emph{typical duration} of threads belonging to each regime.

The mixture weight $w_1 \approx 0.41$--$0.50$ indicates that, across all communities, approximately half of all threads belong to each regime. This roughly equal split suggests that the two temporal regimes are not niche phenomena but fundamental organizational principles of online discussions.

\begin{figure}[htbp]
\centering
\includegraphics[width=\textwidth]{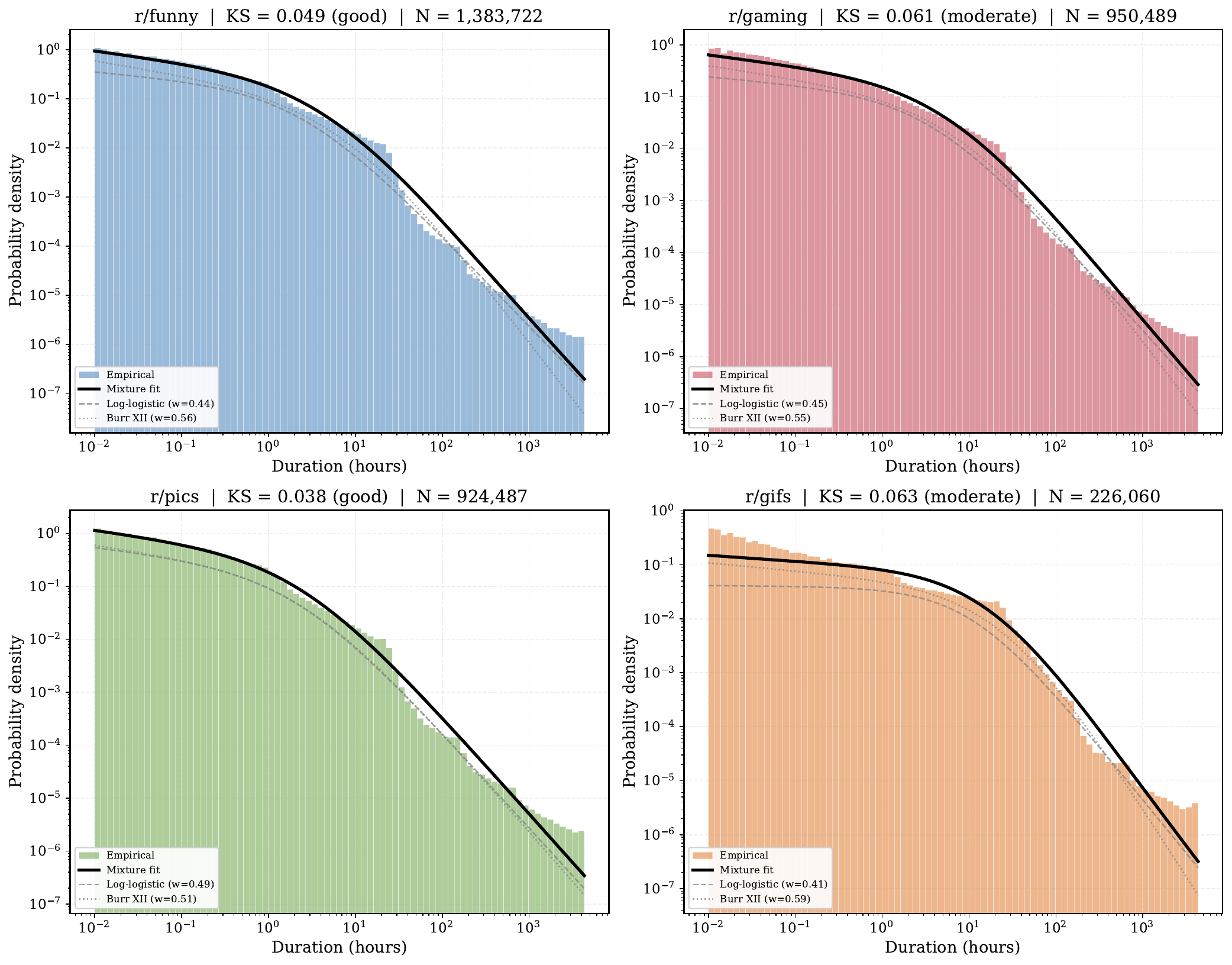}
\caption{}Empirical histograms and fitted mixture densities for the four Reddit communities. Each panel shows the empirical distribution of thread durations (in hours, log scale) as a histogram, overlaid with the fitted two-component Log-Logistic + Burr XII mixture density (solid black line) and the individual component densities (dashed: Log-Logistic component; dotted: Burr XII component). The Log-Logistic component captures shorter-scale interactions (median 2--11 hours), while the Burr XII component captures longer-scale discussions (median 3--26 hours). The mixture model provides an excellent description of the data across all communities, with KS statistics ranging from 0.038 (r/pics) to 0.063 (r/gifs). The vertical axis is scaled to unit area; the horizontal axis is log-scaled in hours.
\label{fig:histograms}
\end{figure}

\subsection{Single Distributions Are Inadequate}

Table~\ref{tab:comparison} compares the best single distributions against the two-component mixture for r/funny. The mixture outperforms all single distributions by a wide margin ($\Delta\mathrm{AICc} = 35{,}164$ over the best single distribution, Burr XII; $\Delta\mathrm{BIC} = 35{,}025$). The power law (Pareto) is among the worst-fitting models ($D = 0.287$), overestimating the probability of extremely long threads while underestimating the bulk of the distribution.

The BIC confirms the AICc results: the mixture model is strongly preferred over all single distributions ($\Delta\mathrm{BIC} > 35{,}000$ for r/funny, and similarly large values for other communities). Because BIC imposes a stronger penalty for additional parameters than AICc (especially given our large sample sizes), this provides particularly strong evidence that the two-component structure is not an artifact of overfitting but reflects genuine structure in the data.

\begin{table}[htbp]
\centering
\caption{Model comparison for r/funny.}
\label{tab:comparison}
\begin{tabular}{lrrrr}
\toprule
Model & Params. & AICc & BIC & KS \\
\midrule
\textbf{Mixture (Log-Logistic + Burr XII)} & \textbf{7} & \textbf{1,117,236} & \textbf{1,117,420} & \textbf{0.049} \\
Burr XII (single)          & 3 & 1,152,400 & 1,152,445 & 0.095 \\
Log-Normal (single)        & 2 & 1,206,400 & 1,206,420 & 0.104 \\
Weibull (single)           & 2 & 1,351,200 & 1,351,220 & 0.178 \\
Power Law (single)         & 2 & 1,629,200 & 1,629,220 & 0.287 \\
Exponential (single)       & 1 & 3,007,200 & 3,007,210 & 0.342 \\
\bottomrule
\end{tabular}
\end{table}

We report detailed results for r/funny as a representative case, the qualitative pattern is identical across communities: in each case, the two-component mixture is decisively preferred (all $\Delta\mathrm{AICc} > 35{,}000$, all $\Delta\mathrm{BIC} > 35{,}000$), the Burr XII is the best single distribution, and the power law ranks among the worst models (KS $> 0.28$ in all communities). The consistency of these results across content types---from humor to gaming to image sharing---indicates that the inadequacy of single distributions is not specific to any one community but reflects a general property of online discussion dynamics.

\subsection{Community Variation in Scale Parameters}

\begin{figure}[htbp]
\centering
\includegraphics[width=0.95\textwidth]{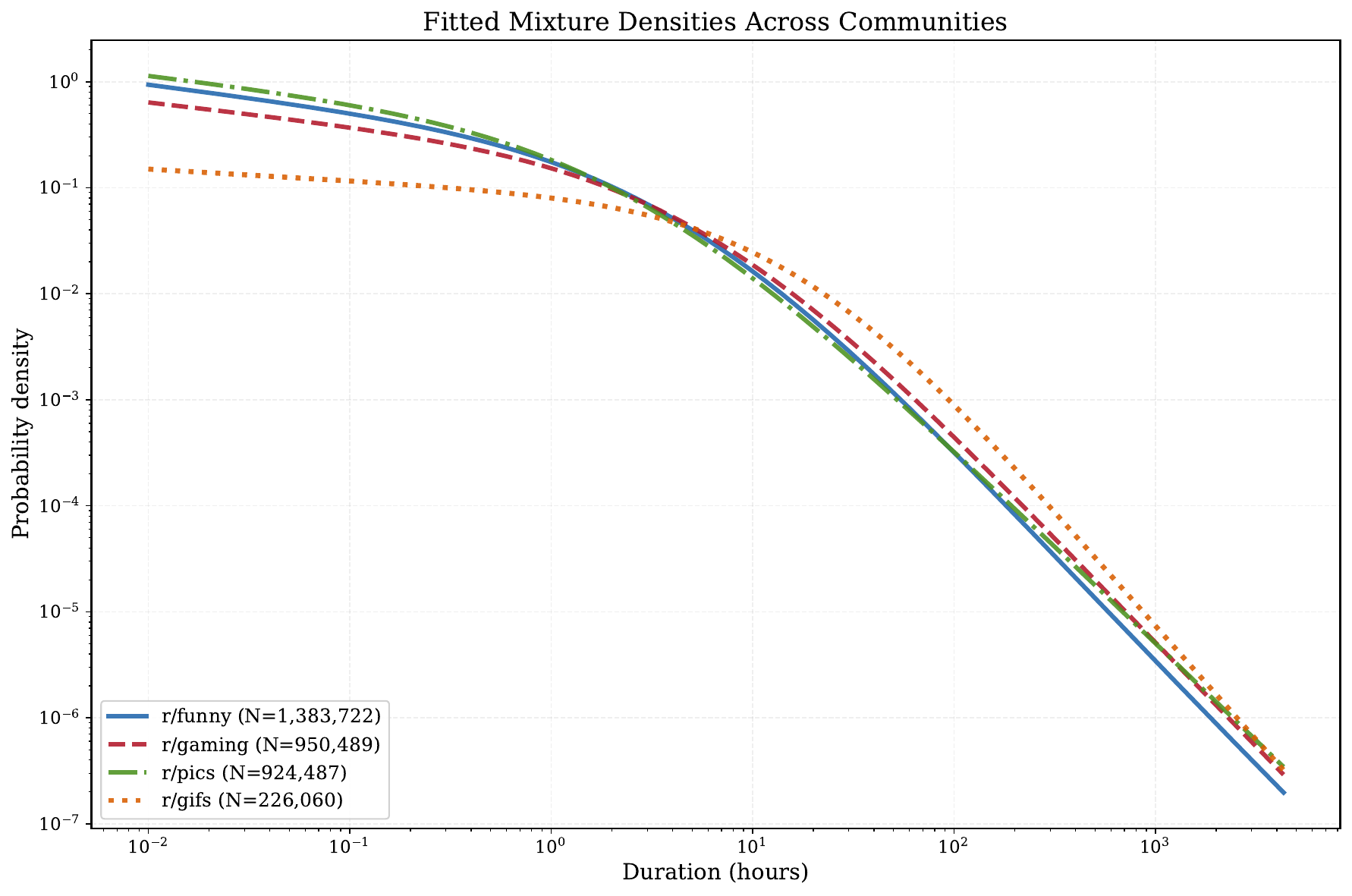}
\caption{Fitted two-component Log-Logistic + Burr XII mixture densities for all four communities. The Log-Logistic component captures shorter-scale interactions, while the Burr XII component captures longer-scale discussions. The mixture weight $w_1$ ranges from 0.41 to 0.50 across communities, indicating a roughly equal split between the two temporal regimes. Community-specific differences in the position and spread of the components reflect variations in interaction rhythms across content types.}
\label{fig:density_comparison}
\end{figure}

\begin{figure}[htbp]
\centering
\includegraphics[width=0.85\textwidth]{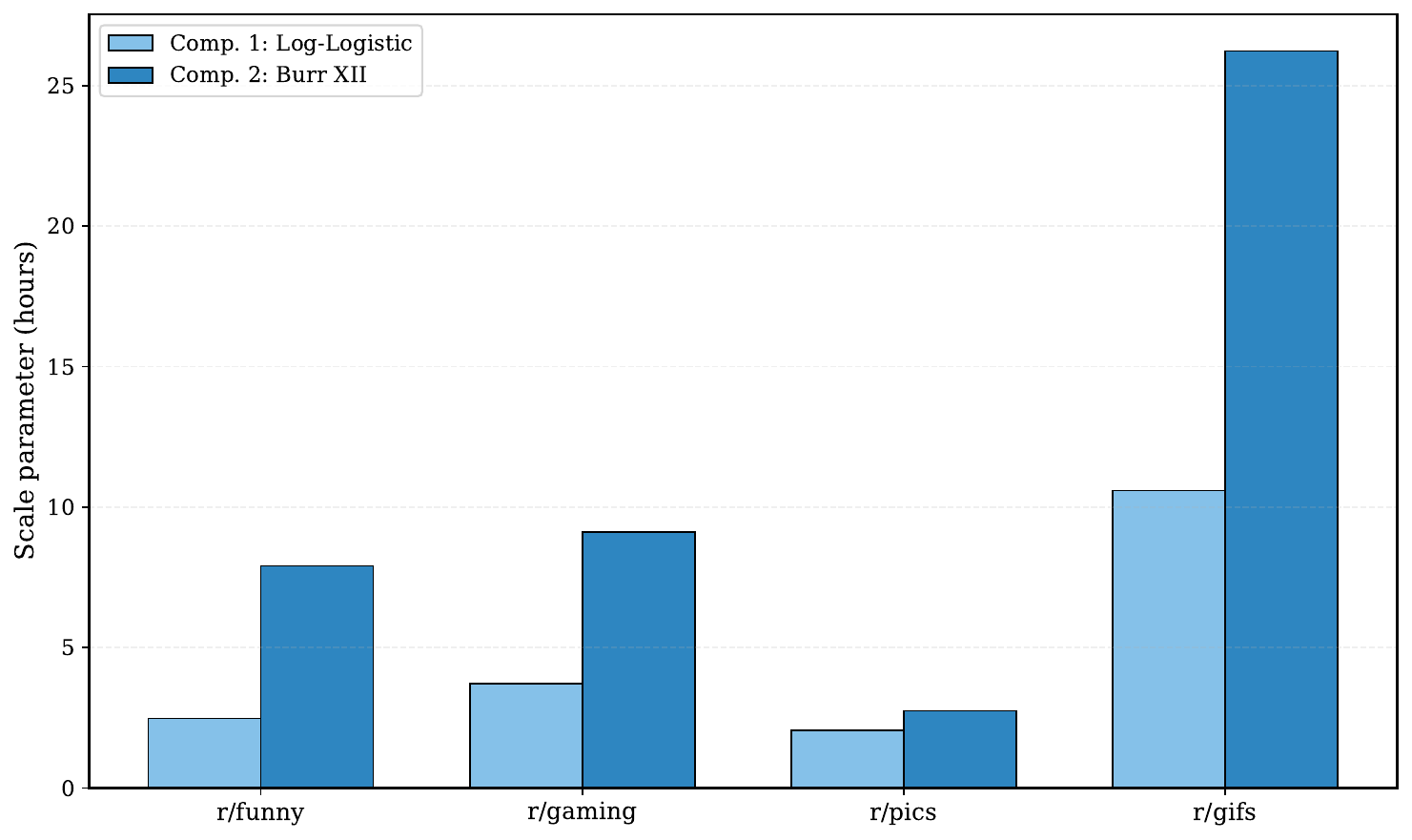}
\caption{Scale parameters (in hours) of the two mixture components across communities. Light bars: Log-Logistic component (scale$_1$); dark bars: Burr XII component (scale$_2$). The Burr XII scale varies by nearly an order of magnitude across communities, from 2.7 hours in r/pics to 26.2 hours in r/gifs, while the Log-Logistic scale ranges from 2.1 to 10.6 hours. These differences indicate that communities are distinguished not by the presence of two temporal regimes, but by the characteristic duration of each regime.}
\label{fig:scale_comparison}
\end{figure}

While the two-component mixture structure is consistent across communities, the scale parameters reveal substantial community-specific variation. Figure~\ref{fig:density_comparison} overlays the fitted mixture densities for all four communities, and Figure~\ref{fig:scale_comparison} quantifies the differences in scale parameters.

The Burr XII scale parameter varies by nearly an order of magnitude across communities (Figure~\ref{fig:scale_comparison}): from 9,897\,s (2.7\,h, r/pics) to 94,429\,s (26.2\,h, r/gifs). This variation reflects the different temporal rhythms of each community. Content-sharing communities like r/pics, where interactions are predominantly brief reactions to images, exhibit the shortest characteristic durations in both components. In contrast, r/gifs shows the largest scale parameters, suggesting that GIF-based exchanges sustain longer interaction cycles---possibly because animated content elicits extended chains of reactions and responses. Discussion-oriented r/gaming and humor-focused r/funny occupy intermediate positions. These differences align with qualitative expectations about community function~\cite{baym2010,kraut2012}, but quantify them through the lens of statistical modeling: communities are distinguished not by whether they exhibit two temporal regimes, but by how long each regime typically lasts.

\subsection{Model Validation}

\begin{figure}[htbp]
\centering\includegraphics[width=\textwidth]{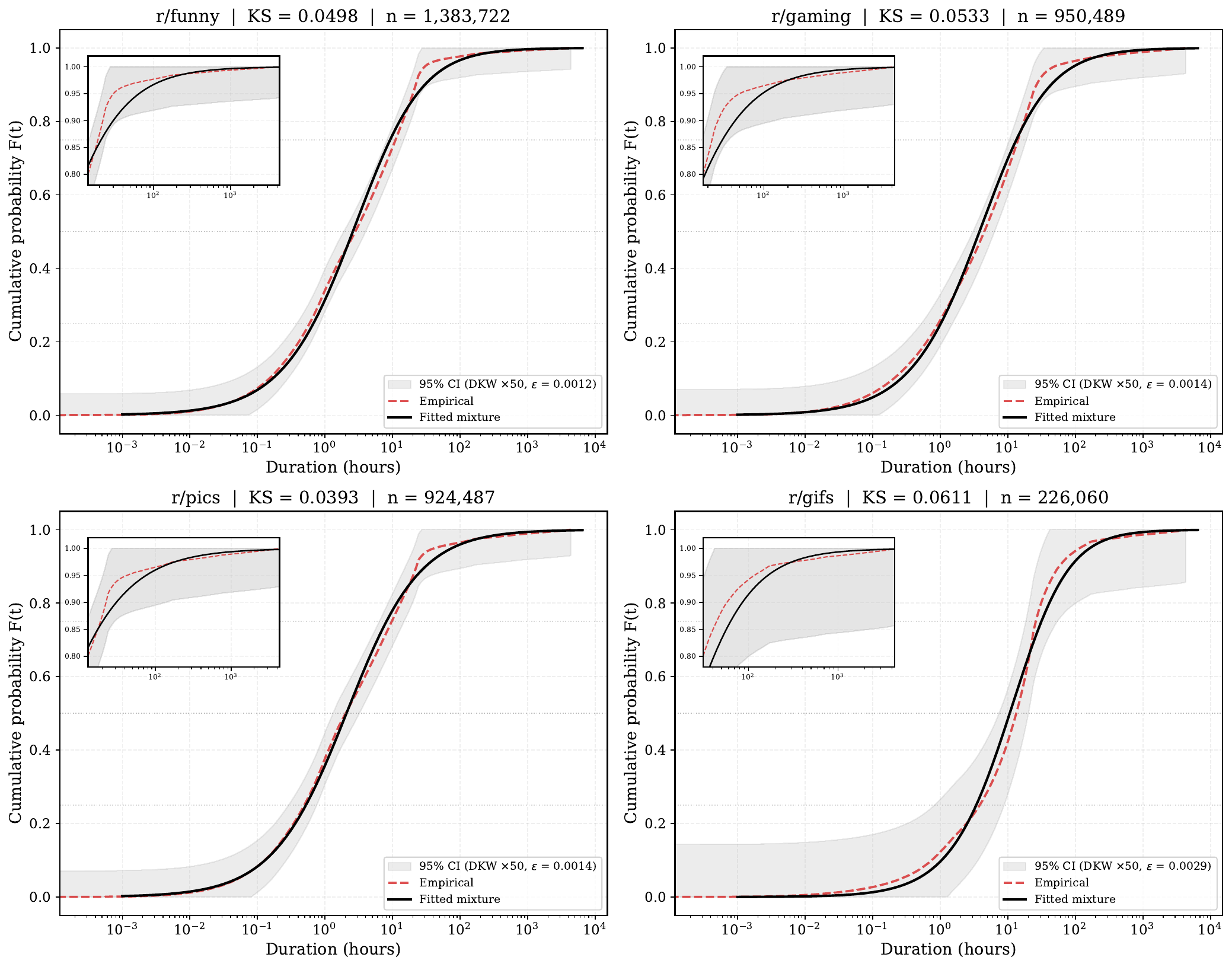}
\caption{Empirical cumulative distribution functions (red dashed lines) versus fitted two-component Log-Logistic + Burr XII mixture CDFs (solid black lines) for all four communities. The close agreement between empirical and fitted curves across the full range of durations demonstrates that the mixture model accurately captures the distribution of thread durations. Insets show a zoom of the upper tail (80th--100th percentiles), where the empirical CDF becomes noisier due to the smaller number of very long threads; the fitted CDF tracks the empirical steps closely even in this sparse region, confirming that the two-component structure adequately describes the extreme tail. The horizontal axis is log-scaled in hours.}
\label{fig:cdf}
\end{figure}

We assessed the quality of the mixture fits using both visual diagnostics and quantitative measures. Figure~\ref{fig:cdf} compares the empirical CDFs against the fitted mixture CDFs for all four communities. The agreement is excellent in the body of the distribution, with deviations concentrated in the upper tail (insets), where the smaller number of very long threads makes the empirical CDF noisier. The insets show that even in the tail region (80th--100th percentile), where the empirical CDF becomes sparser and more step-like due to the smaller number of very long threads, the fitted CDF tracks the empirical steps closely.

\begin{figure}[htbp]
\centering
\includegraphics[width=\textwidth]{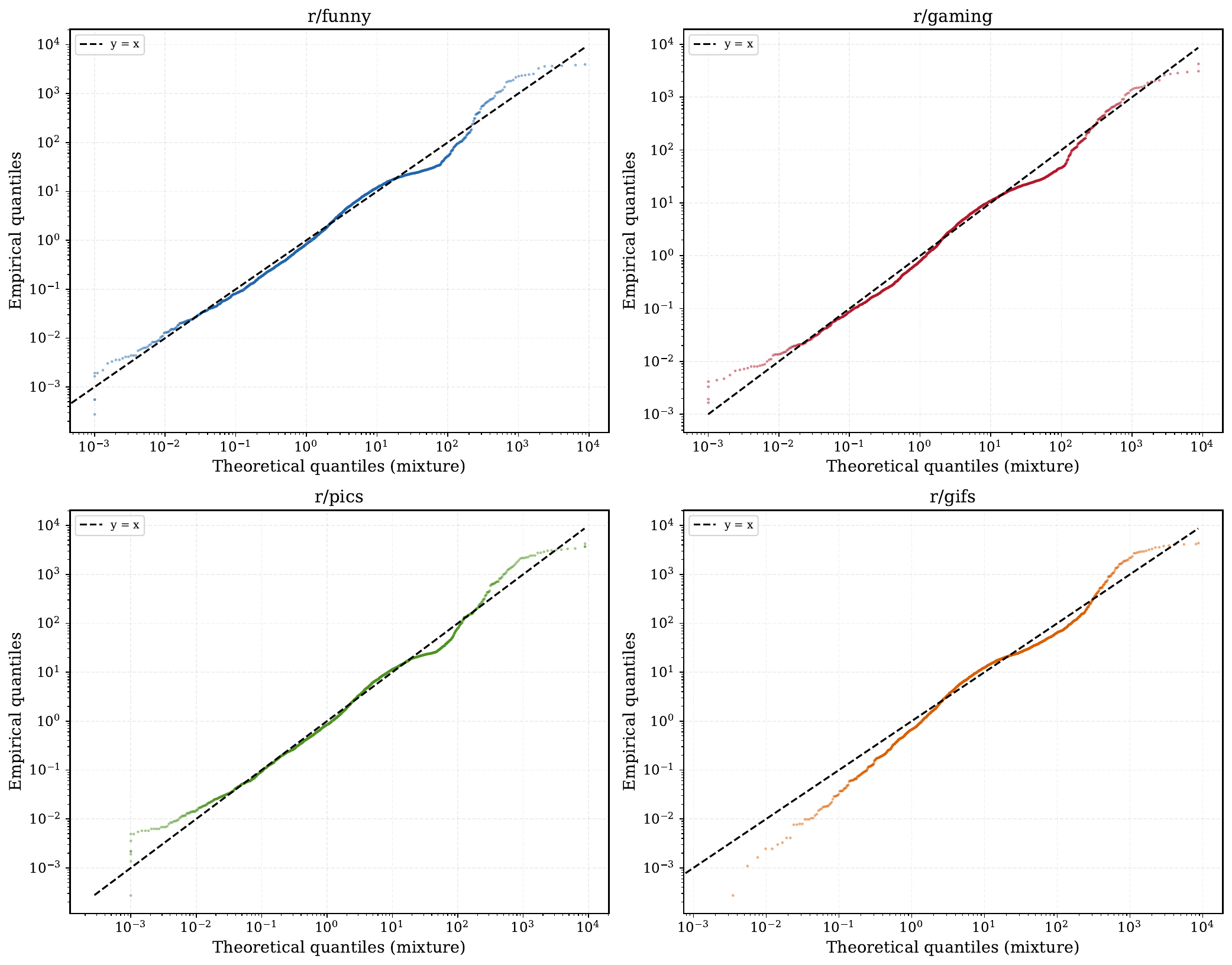}
\caption{Quantile-quantile (QQ) plots comparing empirical quantiles of thread durations against theoretical quantiles from the fitted two-component Log-Logistic + Burr XII mixture for all four communities. Points closely follow the diagonal identity line (red dashed) across the full range of durations, confirming that the mixture model accurately captures the distributional shape from the bulk to the extreme upper tail. Minor deviations are visible only at the very highest quantiles, where the empirical distribution becomes sparser due to the limited number of extremely long threads, and at the very lowest quantiles, where discretization effects and the discrete nature of timestamps (measured in integer seconds) cause slight misalignment. The close alignment of points along the identity line provides strong visual validation of the goodness-of-fit reported by the KS statistics (0.038--0.063).}
\label{fig:qq}
\end{figure}

Figure~\ref{fig:qq} shows QQ plots comparing empirical quantiles against theoretical quantiles from the fitted mixtures. Points closely follow the identity line across the full range, confirming that the mixture model captures the distribution shape accurately, with minor deviations only at the very highest quantiles.

Cross-validation confirms the stability of the model: the coefficient of variation of the log-likelihood across folds is below 0.13\% in all four communities, indicating that the parameter estimates are robust to the choice of training data and that the model does not overfit. The overall quality ratings were ``good'' for r/funny (KS $= 0.049$) and r/pics (KS $= 0.038$), and ``moderate'' for r/gaming (KS $= 0.061$) and r/gifs (KS $= 0.063$). As we show in the temporal analysis below, the moderate fit for r/gifs partly reflects its structural evolution: the model fits substantially better in later years (KS $= 0.054$ in 2020) than in early years (KS $= 0.415$ in 2014).

\subsection{Temporal Robustness}

To assess whether the two-component structure is stable over time or merely an artifact, we fitted the model separately to each year from 2014 to 2020. This temporal analysis serves two purposes: first, it tests whether the mixture structure itself persists across years; second, it reveals whether community interaction patterns are evolving.

\begin{figure}[htbp]
\centering
\includegraphics[width=\textwidth]{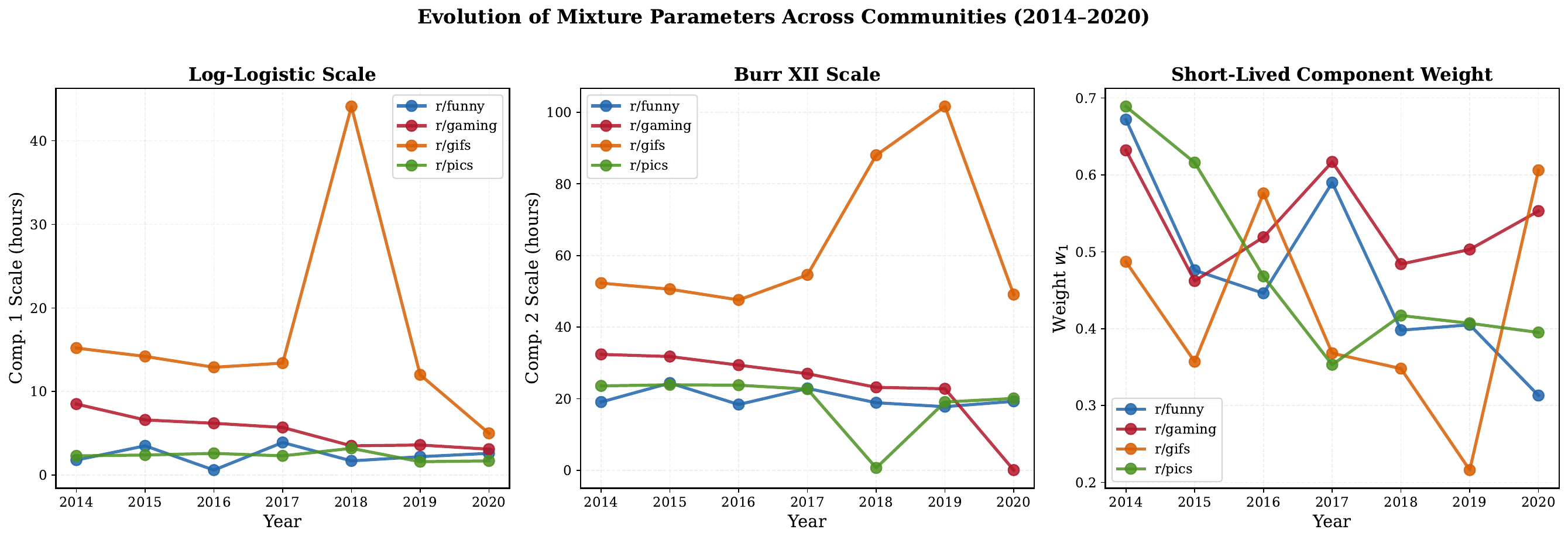}
\caption{Evolution of mixture parameters (2014--2020) across communities. The two-component structure persists across all years, while parameter values drift in community-specific ways. r/funny and r/pics shift toward longer-scale discussions (decreasing $w_1$); r/gaming shows systematic shortening of both scale parameters; and r/gifs undergoes structural maturation (KS improving from 0.415 to 0.054).}
\label{fig:temporal}
\end{figure}

Figure~\ref{fig:temporal} shows the evolution of the three main parameters. The mixture structure remained qualitatively stable across all communities---in every year, the best model was a Log-Logistic $+$ Burr XII mixture, and the two components retained their interpretation as shorter-scale and longer-scale regimes.  

However, while the structure persisted, the parameter values drifted in community-specific ways, revealing three distinct evolutionary patterns. To assess the statistical significance of these trends, we performed linear regression of each parameter against year; the reported p-values correspond to the slope coefficient being different from zero.

\textbf{r/funny and r/pics: shift toward extended discussions.} The weight of the Log-Logistic component decreased significantly (r/funny: $-0.045$/year, $p = 0.031$; r/pics: $-0.048$/year, $p = 0.021$), indicating that a growing fraction of threads belong to the Burr XII regime. These communities, both oriented toward content consumption and humor, appear to be evolving toward more sustained conversations, possibly reflecting changes in user behavior or platform algorithms.

\textbf{r/gaming: systematic shortening.} Both scale parameters decreased significantly (scale$_1$: $-0.89$\,h/year, $p < 0.001$; scale$_2$: $-4.31$\,h/year, $p = 0.018$), indicating that threads are becoming shorter in both regimes. This is consistent with a community becoming more fast-paced, perhaps due to an increasing volume of content competing for attention. Notably, this trend aligns with the conditional analysis finding that user count and thread duration are positively associated: if r/gaming's user base grew over this period, the shortening of threads suggests that other factors (e.g., algorithmic changes, community norms) outweighed the participation effect.

\textbf{r/gifs: structural maturation.} The goodness-of-fit improved markedly (KS: $-0.065$/year, $p = 0.002$), converging from poor fit in early years (KS $= 0.415$ in 2014) to good fit in later years (KS $= 0.054$ in 2020). This suggests that r/gifs, the smallest community in our sample, underwent a process of structural maturation: its temporal dynamics converged toward the two-regime organization observed in the more established communities.

These temporal trends demonstrate that the two-component mixture is not merely a description but a structure that persists over time, even as communities evolve along different trajectories. The emergence of the two-component structure at both the aggregate level (2014--2020) and within individual years suggests a form of \emph{temporal scale invariance}: the distinction between shorter-scale and longer-scale regimes is not an artifact of pooling heterogeneous time periods, but a robust property that manifests consistently regardless of the temporal window chosen. This is consistent with the interpretation that two distinct generative mechanisms---brief reactions and extended discussions---operate at all time scales, and that year-to-year variations reflect changes in the relative prevalence and characteristic duration of these mechanisms (parameter drift) rather than changes in the underlying two-mechanism structure itself. The stability of the qualitative structure---combined with the interpretability of the quantitative drifts---strengthens the case for the mixture model as a robust tool for characterizing online discussion dynamics.

\subsection{Conditional Analysis: Drivers of Mixture Parameters}

\begin{figure}[htbp]
\centering
\includegraphics[width=\textwidth]{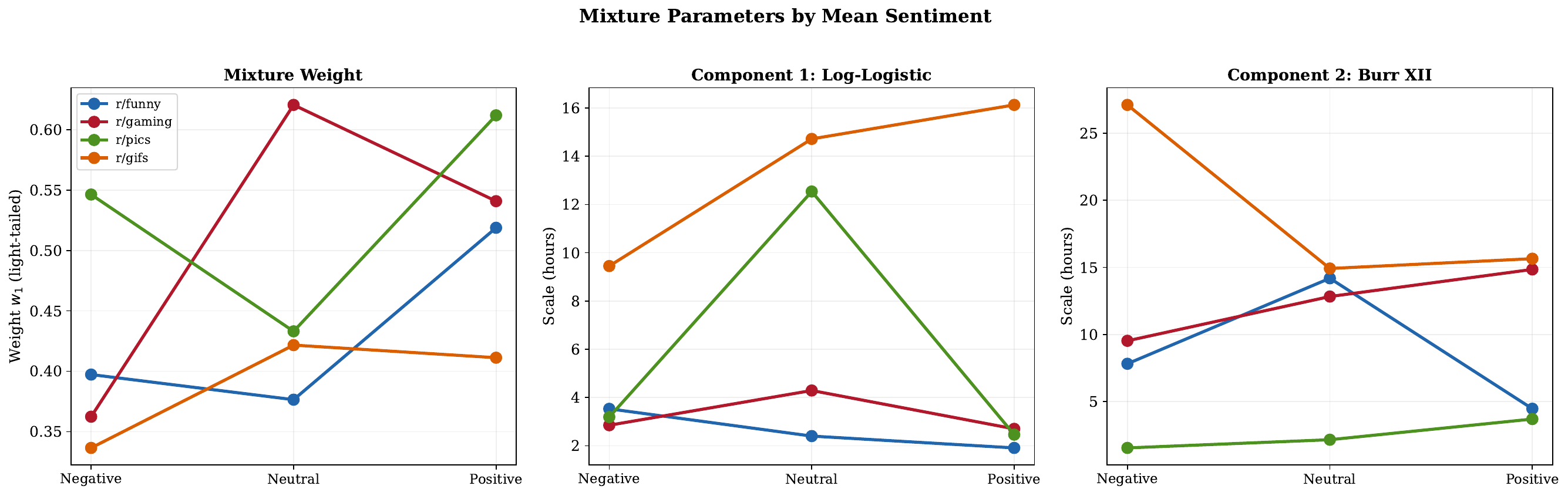}
\caption{Evolution of mixture parameters by mean sentiment class.}
\label{fig:conditional}
\end{figure}

The results presented so far establish that the two-component mixture is a robust statistical description of thread duration distributions, but they do not explain \emph{why} the mixture parameters take the values they do. To investigate this, we stratified threads by three statistically independent variables---number of users, mean sentiment, and conflict polarity---and fitted the Log-Logistic $+$ Burr XII mixture separately to each group. If the mixture parameters vary systematically with thread characteristics, this would indicate that the two temporal regimes are not merely a statistical construct but reflect genuine differences in discussion dynamics driven by participation, emotional tone, and disagreement. Of the 40 groups examined, 35 successfully converged (87.5\%). Table~\ref{tab:conditional} summarizes the qualitative patterns, and Figure~\ref{fig:conditional} shows the evolution of mixture parameters across sentiment groups.

\begin{table}[htbp]
\centering
\caption{Summary of conditional mixture analysis. Arrows indicate direction of change as the stratifying variable increases.}
\label{tab:conditional} 
\small 
\begin{tabular}{lcccc} 
\toprule
\textbf{Variable} & \textbf{$w_1$} & \textbf{scale$_1$} & \textbf{scale$_2$} & \textbf{Interpretation} \\
 & \textbf{(Log-Logistic)} & \textbf{(Log-Logistic)} & \textbf{(Burr XII)} & \\
\midrule
$n_{\text{users}}$ $\uparrow$ & $\approx$ stable & $\uparrow\uparrow$ strong & $\uparrow$ moderate & Both regimes lengthen \\
\addlinespace
$\bar{S}$ $\uparrow$ & $\uparrow$ moderate & $\downarrow$ moderate & $\cap$ peak at $\bar{S}_0$ & Longest durations at neutral \\
\addlinespace
$P_{\text{conf}}$ $\uparrow$ & $\approx$ stable & $\uparrow$ moderate & $\uparrow\uparrow$ strong & Conflict extends longer-scale regime \\
\bottomrule
\end{tabular}
\end{table}

\subsubsection{Number of Users}

Increasing the number of distinct users systematically increased the scale of the Log-Logistic component (scale$_1$) across all four communities. In r/funny, scale$_1$ rose from 4.0\,h (2--5 users) to 22.3\,h (20+ users); in r/pics, from 1.9\,h to 22.7\,h; in r/gaming, from 3.8\,h to 24.9\,h; and in r/gifs, from 10.2\,h to 23.9\,h. This indicates that even threads belonging to the Log-Logistic regime---those that do not develop into extended discussions---become substantially longer when more users participate. The weight $w_1$ remained approximately stable across user groups, suggesting that the number of participants does not determine whether a thread becomes a brief reaction or an extended discussion, but rather lengthens the characteristic duration within whichever regime the thread belongs to.

This finding has a straightforward interpretation: more participants generate more comments, which takes more time, regardless of the nature of the discussion. The effect is most pronounced for the Log-Logistic component, where the relative increase in scale is largest (e.g., a factor of 5.6 in r/funny), and more moderate for the Burr XII component. This asymmetry suggests that participation breadth is especially important for prolonging interactions that would otherwise remain brief, whereas extended discussions are sustained by additional mechanisms beyond mere user count.

\subsubsection{Mean Sentiment}

Neutral threads ($\bar{S}_{0}$) exhibited the lowest weight on the Log-Logistic component ($w_1$) in three of four communities (r/funny: 0.376 vs.\ 0.397 for $\bar{S}_{-}$ and 0.519 for $\bar{S}_{+}$; r/pics: 0.433 vs.\ 0.546 and 0.612; r/gifs: 0.422 vs.\ 0.337 and 0.411). This indicates that neutral threads are less likely to belong to the Log-Logistic regime---they have a higher probability of developing into extended discussions.

More strikingly, neutral threads exhibited the largest scale$_2$ among the three sentiment classes in three of the four communities. In r/funny, the Burr XII scale for $\bar{S}_{0}$ (14.2\,h) was nearly twice that of $\bar{S}_{-}$ (7.8\,h) and more than three times that of $\bar{S}_{+}$ (4.5\,h). The same pattern held in r/pics ($\bar{S}_{0}$: 2.2\,h vs.\ $\bar{S}_{-}$: 1.5\,h vs.\ $\bar{S}_{+}$: 3.7\,h) and r/gifs ($\bar{S}_{0}$: 14.9\,h vs.\ $\bar{S}_{-}$: 27.1\,h vs.\ $\bar{S}_{+}$: 15.6\,h). In r/gaming, however, the largest scale$_2$ was observed for positive sentiment (14.9\,h), slightly exceeding the neutral value (12.8\,h). Overall, neutral sentiment is consistently associated with either the largest or near-largest scale parameters across communities, confirming its link to prolonged engagement.

Neutral sentiment is thus associated with both a higher probability of entering the Burr XII regime and, in most communities, substantially heavier tails once in that regime.

This result might suggest that neutral-toned discussions tend to persist longer than emotionally charged ones. However, as we show in Section 3.8, this "neutral advantage" is largely explained by the fact that neutral threads are mostly high-conflict threads, with evenly balanced opposing factions maintaining conflict for extended periods.

\subsubsection{Conflict Polarity}

Higher conflict polarity systematically increased the scale of the Burr XII component across all four communities. In r/funny, scale$_2$ rose from 1.9\,h (no conflict) to 7.9\,h (high conflict); in r/gaming, from 5.6\,h to 8.9\,h; in r/pics, from 1.5\,h to 2.9\,h; and in r/gifs, from 1.6\,h to 24.9\,h. The weight $w_1$ showed no systematic trend across conflict levels, remaining stable in three communities while increasing moderately in r/pics (from 0.426 to 0.568), indicating that conflict polarity does not determine whether a thread enters the Burr XII regime, but rather how long it persists once it does.

This result aligns with and extends prior work showing that controversy increases engagement~\cite{chen2013}. Our distributional analysis reveals that the effect of conflict is specifically concentrated in the upper tail of the duration distribution: high-conflict threads do not simply last longer on average, but exhibit disproportionately extended durations, consistent with the idea that opposing factions sustain engagement by repeatedly returning to the discussion. The effect is particularly pronounced in communities with inherently longer interaction cycles (r/gifs: scale$_2$ increases by a factor of 15.6 from no conflict to high conflict), suggesting that conflict amplifies existing community-specific temporal tendencies rather than imposing a uniform effect across contexts.

\subsubsection{Community Consistency}

The qualitative patterns described above were found to be consistent across three or four of the four communities analyzed. The effect of user count on the scale$_1$ was observed in all four communities; the lower weight of the Log-Logistic component for neutral threads was found in three of the four communities (r/funny, r/pics, and r/gifs, with r/gaming as the exception where neutral threads exhibited the highest $w_1$); and the tail-lengthening effect due to conflict polarity was observed in all four communities. This replication across communities is noteworthy, given the diversity of the communities analyzed, which have different user bases and norms. The fact that similar relationships between thread characteristics and mixture parameters emerge in these different contexts suggests that they reflect fundamental properties of online discussions rather than community-specific characteristics.

At the same time, the magnitude of these effects varies across communities in ways that align with qualitative expectations. The conflict effect on scale$_2$ is strongest in r/gifs (15.6$\times$) and weakest in r/gaming (1.6$\times$), with r/pics (1.9$\times$) and r/funny (4.2$\times$) occupying intermediate positions. This variation is consistent with the observation that communities with inherently longer interaction cycles, such as r/gifs, exhibit greater amplification of duration through conflict, while communities with shorter baseline durations, such as r/pics, show more modest effects. This variation in effect size, combined with consistency in the effect's direction, provides a nuanced picture: the structural factors that determine thread duration are universal, but their impact is modulated by community context.

\subsection{The Neutral Sentiment--Conflict Polarity Relationship}

\begin{figure}[htbp]
\centering
\includegraphics[width=\textwidth]{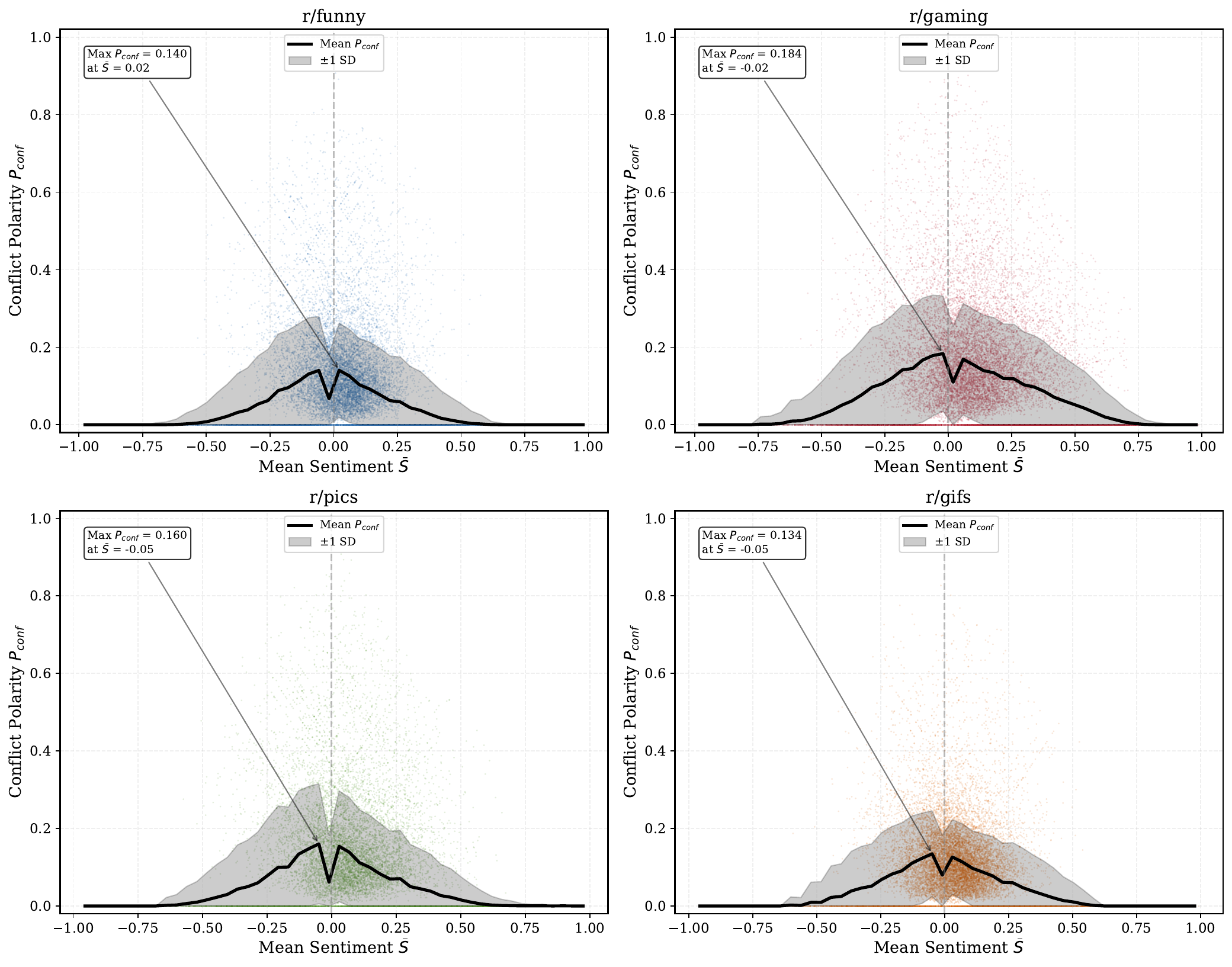}
\caption{Conflict polarity ($P_{\text{conf}}$) vs mean sentiment ($\bar{S}$). Each panel shows a random sample of 20,000 threads (colored points), the binned mean $P_{\text{conf}}$ (black line), and $\pm 1$ standard deviation (gray band). The vertical dashed line marks $\bar{S} = 0$. In all four communities, $P_{\text{conf}}$ peaks within the neutral sentiment interval ($\bar{S} \in [-0.2, 0.2]$).}
\label{fig:sentiment_conflict}
\end{figure}

The finding that neutral threads exhibit heavier tails raises a question: is neutral sentiment itself beneficial for thread longevity, or is it a proxy for another variable? Figure~\ref{fig:sentiment_conflict} plots conflict polarity against mean sentiment for all four communities. Across all four communities, the relationship exhibits a clear inverted U-shape: conflict polarity peaks at $\bar{S} \approx 0$ and decreases toward both extremes of the sentiment scale. The maximum mean $P_{\text{conf}}$ falls within the neutral stratification interval ($\bar{S} \in [-0.2, 0.2]$) in every community. This reveals that threads classified as ``neutral'' by mean sentiment are not genuinely neutral---they are threads with \emph{balanced opposing factions}: approximately half of the users express positive sentiment and half express negative sentiment, yielding a mean near zero but a high conflict polarity.

This finding resolves the apparent longevity of neutral threads. In r/funny, the heavier tails and lower Log-Logistic weight observed for $\bar{S}_{0}$ threads are clearly attributable to the high conflict polarity that peaks in the neutral interval. However, the relationship is more nuanced across communities: in r/gaming, neutral threads exhibit the highest $w_1$ (0.621), indicating that they are more likely to belong to the shorter-scale regime despite their high conflict polarity, while in r/pics and r/gifs, the tails of neutral threads are not uniformly the heaviest among sentiment classes. Nevertheless, the overarching pattern is that mean sentiment acts as a proxy for conflict polarity: threads with $\bar{S} \approx 0$ are disproportionately high-conflict threads, and it is the conflict---not the neutrality per se---that sustains extended discussions. This interpretation is consistent with the conditional analysis results for $P_{\text{conf}}$ (Section~3.7.3), where higher conflict polarity systematically increased scale$_2$ across all communities.

This also explains why the pairwise correlation between $\bar{S}$ and $P_{\text{conf}}$ is low despite their strong link: the relationship is symmetric and non-monotonic (inverted U-shaped), which linear correlation cannot capture. The stratification into three sentiment classes partially recovers this nonlinearity, but the continuous plot in Figure~\ref{fig:sentiment_conflict} provides the full picture.

\section{Discussion}

Our results paint a consistent picture of online discussion dynamics. Across 3.5 million threads from four diverse Reddit communities, thread durations are best described not by a single distribution but by a mixture of two components, reflecting the coexistence of brief reactions (shorter-scale) and extended conversations (longer-scale). In this section, we first address the evidence supporting the choice of two components, then discuss the limitations of power-law models, which we find to be among the worst-fitting distributions. We next examine the structural drivers of thread duration---participation breadth, sentiment, and conflict polarity---and show how the apparent longevity of neutral-tone discussions is explained by their disproportionate conflict polarity. We then consider the community-specific variation in scale parameters, which span nearly an order of magnitude, and the temporal evolution of these patterns over the 2014--2020 period. Finally, we discuss the implications of our findings for understanding online engagement, content recommendation, and generative models of collective behavior.

\subsection{The Two-Component Mixture as a Robust Approximation}
Across the four Reddit communities we analyzed, thread durations are consistently best described by a two-component mixture of Log-Logistic and Burr XII distributions. While we cannot claim universality on the basis of four communities from a single platform, several lines of evidence support the robustness of this characterization.

The choice of two mixture components is supported by multiple lines of evidence. First, the AICc improvement from the best single distribution (Burr XII) to the two-component mixture is substantial ($\Delta\mathrm{AICc} > 35{,}000$ across all communities), and the BIC confirms this preference ($\Delta\mathrm{BIC} > 35{,}000$), confirming that a single distribution is insufficient even under the more conservative BIC penalty. Second, 5-fold cross-validation shows that the two-component model is stable (CV $< 0.13\%$), indicating no overfitting despite the additional parameters. Third, 35 of 40 conditional fits (87.5\%) converged successfully, demonstrating that the two-component structure is robust across different thread characteristics. The five groups that failed to converge were systematically in intermediate regimes where the two components overlap most heavily: the 5--20 user group in r/funny, and the low and medium conflict polarity groups in r/funny and r/pics. No failures occurred at the extremes (e.g., 2--5 users, high conflict), where the Log-Logistic and Burr XII components are well separated. Fourth, we attempted to fit mixtures with three components on r/funny using Nelder-Mead optimization with up to 50 restarts; these fits failed to converge, producing infinite AICc values due to parameter explosion. While a definitive test using EM across all communities remains future work, the failure of K $\geq$ 3 under direct optimization, combined with the excellent fit of K = 2 (KS = 0.038 for r/pics, 0.049 for r/funny, 0.061 for r/gaming, and 0.063 for r/gifs) and the systematic pattern of fit failures in intermediate groups, strongly suggests that two components capture the essential structure of the data. Finally, the temporal analysis shows that the two-component structure persists across seven years of data (2014--2020), with community-specific parameter drifts rather than structural changes. The mixture weight $w_1$ remains consistently in the range 0.41--0.50 across communities, further supporting the stability of the two-regime organization. Taken together, these results indicate that two components provide a parsimonious and robust description of thread duration distributions, capturing the essential distinction between shorter-scale (brief reactions) and longer-scale (extended discussions) regimes.

\subsection{Why Power Laws Fail}

Despite the popularity of power laws in online media research~\cite{newman2005}, our analysis strongly rejects the Pareto distribution for thread durations across all four communities. The KS statistic for the Pareto distribution exceeds 0.28 in every community (e.g., $D = 0.287$ for r/funny, and similarly large values for the others), indicating a poor fit that systematically overestimates the probability of extremely long threads while underestimating the bulk of the distribution. The Burr XII, which includes the Pareto as a limiting case~\cite{kleiber2003}, provides a substantially better single-distribution fit ($D = 0.095$ for r/funny, and KS values below 0.10 in all communities) because its additional shape parameters capture the curvature in the empirical distribution that a pure power law cannot. However, even the Burr XII is far outperformed by the two-component mixture ($D = 0.038$--$0.063$, $\Delta\mathrm{AICc} > 35{,}000$), demonstrating that no single distribution---whether power law or otherwise---adequately captures the full complexity of thread duration dynamics. The log-normal distribution, often proposed as an alternative to power laws~\cite{mitzenmacher2004}, also provides a better fit than the Pareto ($D = 0.104$ for r/funny) but remains substantially worse than the Burr XII and the mixture. These results are consistent with the methodological caution described in~\cite{clauset2009} regarding the importance of rigorous testing against flexible alternatives, and with the large-scale analysis in~\cite{broido2019} questioning the prevalence of power laws in real networks. Our findings add to this growing body of evidence by demonstrating that even when a flexible heavy-tailed distribution like the Burr XII is considered, a single-component model remains insufficient---the two-regime structure is essential.

\subsection{Structural Drivers of Duration}

The conditional analysis reveals that mixture parameters respond systematically to thread characteristics. More users lengthen both duration regimes across all four communities, with the effect being most pronounced for the Log-Logistic component (scale$_1$ increases by factors ranging from 2.3$\times$ in r/gifs to 11.9$\times$ in r/pics). Conflict polarity consistently extends the Burr XII component (scale$_2$) across all communities, with the strongest effect in r/gifs (15.6$\times$) and the weakest in r/gaming (1.6$\times$).

The effect of mean sentiment on the mixture parameters is more nuanced. In r/funny, neutral threads exhibit lower $w_1$ and heavier tails, consistent with the interpretation that conflict drives longevity. However, the relationship varies across communities: in r/gaming, neutral threads exhibit the highest $w_1$ (0.621), while positive sentiment shows the largest scale$_2$ (14.9\,h); in r/pics and r/gifs, neutral threads do not uniformly exhibit the heaviest tails. Despite these variations, the overarching pattern is that neutral threads are disproportionately high-conflict threads, as revealed by the U-shaped relationship between $\bar{S}$ and $P_{\text{conf}}$ (Figure~\ref{fig:sentiment_conflict}), and conflict, not neutrality per se, is the primary driver of extended discussions.

This finding reconciles the conditional analysis results with the pairwise independence of $\bar{S}$ and $P_{\text{conf}}$. The relationship is nonlinear and symmetric, invisible to linear correlation but clearly revealed by the conditional stratification and the continuous sentiment--conflict plot. It also provides an interpretation for the longevity of neutral threads reported in prior survival analyses of online discussions: neutral sentiment is a marker for balanced controversy, which sustains engagement over extended periods.

The structural drivers we identify (participation, conflict) are not static but evolve over time. The shortening of threads in r/gaming despite increasing user counts suggests that platform-level factors can override the participation effect, and the convergence of r/gifs toward the two-regime structure---as evidenced by the improving KS values from 0.415 in 2014 to 0.054 in 2020---suggests a maturation process that may apply to other emerging communities. These temporal dynamics highlight that while the two-regime structure is robust, the relative prevalence and characteristic duration of each regime are modulated by both community context and evolutionary changes over time.

\subsection{Limitations and Future Work}

Several limitations should be noted. First, our analysis is restricted to threads with at least two interactions, excluding single-comment posts; while this is necessary for defining duration, it means our results do not characterize the vast number of posts that receive no replies or only a single reply. Second, our exploratory attempts to fit three-component mixtures with Nelder-Mead optimization failed to converge; a systematic comparison using EM across all communities would be valuable to definitively test whether K $\geq$ 3 provides a significant improvement, though the excellent fit of the two-component model (KS = 0.038--0.063) and the interpretability of the two regimes suggest that additional components may not be necessary. Third, the KS distances for r/gaming (0.061) and r/gifs (0.063) indicate moderate fit, suggesting that additional structure—potentially community-specific nuances or time-varying dynamics—remains uncaptured by our model. Fourth, our sentiment analysis relies on VADER, which, while widely used and validated for social media text, may not fully capture the nuanced emotional content of online discussions, particularly sarcasm or irony. Fifth, our analysis is descriptive rather than generative; models that generate the observed structure from first principles (e.g., self-exciting Hawkes processes~\cite{hawkes1971}, or agent-based models of attention dynamics) would provide deeper understanding. Sixth, our analysis is restricted to four English-language Reddit communities; validation on other platforms (e.g., Twitter, Facebook, or discussion forums) and in other languages is needed to establish the generality of our findings beyond this specific context. Seventh, the temporal analysis covers seven years (2014--2020), but we cannot disentangle the effects of platform algorithm changes, user base evolution, cultural shifts, or external events (e.g., the COVID-19 pandemic) that may have influenced online behavior during this period.

Future work should address these limitations by extending the analysis to additional platforms and languages, incorporating richer sentiment models (e.g., transformer-based approaches), and developing generative models that can reproduce the two-component structure from underlying cognitive or social mechanisms. Additionally, investigating whether the two-regime organization emerges in other types of online interactions—such as email threads, collaborative editing, or real-time chat—would help establish the universality of this temporal structure.

\section{Conclusion}

This study provides a comprehensive statistical characterization of discussion duration across 3.5 million conversations from four Reddit communities. Three main findings emerge.

First, discussion duration is consistently best described by a two-component mixture of Log-Logistic and Burr XII distributions, with approximately half of all discussions belonging to each component. The Log-Logistic component captures shorter-scale, reactive interactions lasting a few hours, while the Burr XII component captures longer-scale discussions that can extend over days or weeks. The widely hypothesized power-law model is among the least suitable models, systematically overestimating the likelihood of extremely long discussions. This finding adds to the growing body of evidence (\cite{clauset2009, broido2019}) demonstrating that power laws should not be assumed without rigorous testing against flexible alternatives. Both AICc and BIC strongly support the mixture model ($\Delta\mathrm{AICc} > 35{,}000$, $\Delta\mathrm{BIC} > 35{,}000$), confirming that the additional parameters are well justified.

Second, the two-component structure proves robust across different community types (humor, games, image sharing, GIFs), across stratification variables (number of users, sentiment, conflict), and over time (stable from 2014 to 2020, despite variations in community-specific parameters). This temporal invariance—the fact that the same structure emerges at both the annual and aggregate levels—suggests that the two regimes reflect genuine and distinct generative mechanisms, rather than statistical artifacts. At the same time, community-specific variation is substantial: the Burr XII scale parameter ranges from 2.7\,h (r/pics) to 26.2\,h (r/gifs), reflecting different temporal rhythms across communities, and temporal trends reveal systematic evolution, such as the shortening of threads in r/gaming and the structural maturation of r/gifs.

Third, the mixture parameters respond systematically to social variables. A greater number of users lengthens the duration of both regimes, with the effect most pronounced for the shorter-scale component. Conflict polarity consistently extends the Burr XII component across all communities: opposing factions maintain interest by repeatedly returning to the debate. The apparent longevity of neutral-tone discussions is explained by their disproportionately high conflict polarity: average neutral sentiment is an indicator of conflict between balanced factions, not a determinant of interaction itself. While this pattern is most clearly observed in r/funny, the relationship between sentiment and duration is more nuanced in other communities—for instance, in r/gaming, positive sentiment exhibits the largest scale parameter—highlighting that community context modulates the expression of these structural drivers.

In addition to establishing a basic model for discussion duration, this study demonstrates that the statistical structure of online conversations is not arbitrary, but reflects underlying social dynamics in predictable and interpretable ways. The two-component mixture provides a compact yet expressive description of how conversations develop in online communities, a description that can be useful for content recommendation, community health monitoring, and generative models of collective behavior. Furthermore, the systematic variation of mixture parameters based on discussion characteristics suggests that distributional analysis can serve as a complementary tool to traditional regression and survival analysis methods for understanding online social systems, offering a distributional perspective that captures the full range of temporal dynamics rather than focusing solely on averages or survival probabilities.

\section*{Acknowledgements}

The computing resources and the related technical support used for this work have been provided by CRESCO/ENEAGRID High Performance Computing infrastructure and its staff~\cite{iannone2019}. CRESCO/ENEAGRID High Performance Computing infrastructure is funded by ENEA, the Italian National Agency for New Technologies, Energy and Sustainable Economic Development and by Italian and European research programmes, see \url{http://www.cresco.enea.it/} for information.

\bibliographystyle{plainnat}

\begin{thebibliography}{99}

\bibitem{madoc}
M.~Mitrovi\'{c}~Dankulov et~al.
\newblock MADOC: Multi-Platform Aggregated Dataset of Online Communities.
\newblock In \textit{Proceedings of the International AAAI Conference on Web and Social Media}, 19(1):2529--2538, 2025.
\newblock doi: https://doi.org/10.1609/icwsm.v19i1.35954

\bibitem{wang2017}
L.~Wang, A.~Ramachandran, and A.~Chaintreau.
\newblock Measuring click and share dynamics on social media.
\newblock In \textit{Proceedings of the ACM Conference on Online Social Networks (COSN)}, 2017.

\bibitem{medvedev2018}
A.~Medvedev, R.~Lambiotte, and J.-C.~Delvenne.
\newblock The anatomy of Reddit: An overview of academic research.
\newblock \textit{arXiv preprint arXiv:1810.10825}, 2018.

\bibitem{lakkaraju2013}
H.~Lakkaraju, J.~McAuley, and J.~Leskovec.
\newblock What's in a name? Understanding the interplay between titles, content, and communities in social media.
\newblock In \textit{Proceedings of the International AAAI Conference on Web and Social Media (ICWSM)}, 7(1):311--320, 2013.
\newblock doi: https://doi.org/10.1609/icwsm.v7i1.14408

\bibitem{szabo2010}
G.~Szabo and B.~A.~Huberman.
\newblock Predicting the popularity of online content.
\newblock \textit{Communications of the ACM}, 53(8):80--88, 2010.
\newblock doi: https://doi.org/10.1145/1787234.1787254

\bibitem{barabasi2005}
A.-L.~Barab\'{a}si.
\newblock The origin of bursts and heavy tails in human dynamics.
\newblock \textit{Nature}, 435(7039):207--211, 2005.
\newblock doi: https://doi.org/10.1038/nature03459

\bibitem{crane2008}
R.~Crane and D.~Sornette.
\newblock Robust dynamic classes revealed by measuring the response function of a social system.
\newblock \textit{Proceedings of the National Academy of Sciences (PNAS)}, 105(41):15649--15653, 2008.
\newblock doi: https://doi.org/10.1073/pnas.0803685105

\bibitem{leskovec2007}
J.~Leskovec, M.~McGlohon, C.~Faloutsos, N.~Glance, and M.~Hurst.
\newblock Patterns of cascading behavior in large blog graphs.
\newblock In \textit{Proceedings of the SIAM International Conference on Data Mining (SDM)}, pages 551--556, 2007.
\newblock doi: https://doi.org/10.1137/1.9781611972771.60

\bibitem{gomez2013}
V.~G\'{o}mez, H.~J.~Kappen, and A.~Kaltenbrunner.
\newblock Modeling the structure and evolution of discussion cascades.
\newblock In \textit{Proceedings of the ACM Conference on Hypertext and Hypermedia (HT)}, pages 181--190, 2011.
\newblock doi: https://doi.org/10.1145/1995966.1995992

\bibitem{cheng2014}
J.~Cheng, L.~A.~Adamic, P.~A.~Dow, J.~Kleinberg, and J.~Leskovec.
\newblock Can cascades be predicted?
\newblock In \textit{Proceedings of the 23rd International Conference on World Wide Web (WWW)}, pages 925--936, 2014.
\newblock doi: https://doi.org/10.1145/2566486.2567997

\bibitem{wu2007}
F.~Wu and B.~A.~Huberman.
\newblock Novelty and collective attention.
\newblock \textit{Proceedings of the National Academy of Sciences (PNAS)}, 104(45):17599--17601, 2007.
\newblock doi: https://doi.org/10.1073/pnas.0704916104

\bibitem{lerman2016}
K.~Lerman.
\newblock Information is not a virus, and other consequences of human cognitive limits.
\newblock \textit{Future Internet}, 8(2):21, 2016.
\newblock doi: https://doi.org/10.3390/fi8020021

\bibitem{aragon2017}
P.~Arag\'{o}n, V.~G\'{o}mez, D.~Garc\'{i}a, and A.~Kaltenbrunner.
\newblock Generative models of online discussion threads.
\newblock \textit{Social Network Analysis and Mining}, 7(1):1--16, 2017.
\newblock doi: https://doi.org/10.1007/s13278-017-0470-3

\bibitem{backstrom2013}
L.~Backstrom, J.~Kleinberg, L.~Lee, and C.~Danescu-Niculescu-Mizil.
\newblock Characterizing and curating conversation threads.
\newblock In \textit{Proceedings of the 6th ACM International Conference on Web Search and Data Mining (WSDM)}, pages 13--22, 2013.
\newblock doi: https://doi.org/10.1145/2433396.2433401

\bibitem{mitzenmacher2004}
M.~Mitzenmacher.
\newblock A brief history of generative models for power law and lognormal distributions.
\newblock \textit{Internet Mathematics}, 1(2):226--251, 2004.
\newblock doi: https://doi.org/10.1080/15427951.2004.10129088

\bibitem{newman2005}
M.~E.~J.~Newman.
\newblock Power laws, Pareto distributions and Zipf's law.
\newblock \textit{Contemporary Physics}, 46(5):323--351, 2005.
\newblock doi: https://doi.org/10.1080/00107510500052444

\bibitem{clauset2009}
A.~Clauset, C.~R.~Shalizi, and M.~E.~J.~Newman.
\newblock Power-law distributions in empirical data.
\newblock \textit{SIAM Review}, 51(4):661--703, 2009.
\newblock doi: https://doi.org/10.1137/070710111

\bibitem{stumpf2012}
M.~P.~H.~Stumpf and M.~A.~Porter.
\newblock Critical truths about power laws.
\newblock \textit{Science}, 335(6069):665--666, 2012.
\newblock doi: https://doi.org/10.1126/science.1216142

\bibitem{broido2019}
A.~D.~Broido and A.~Clauset.
\newblock Scale-free networks are rare.
\newblock \textit{Nature Communications}, 10(1):1017, 2019.
\newblock doi: https://doi.org/10.1038/s41467-019-08746-5

\bibitem{vosoughi2018}
S.~Vosoughi, D.~Roy, and S.~Aral.
\newblock The spread of true and false news online.
\newblock \textit{Science}, 359(6380):1146--1151, 2018.
\newblock doi: https://doi.org/10.1126/science.aap9559

\bibitem{baym2010}
N.~K.~Baym.
\newblock \textit{Personal Connections in the Digital Age}.
\newblock Polity Press, 2010.

\bibitem{kraut2012}
R.~E.~Kraut and P.~Resnick.
\newblock \textit{Building Successful Online Communities: Evidence-Based Social Design}.
\newblock MIT Press, 2012.
\newblock doi: https://doi.org/10.7551/mitpress/9780262016575.001.0001

\bibitem{marsaglia2003}
G.~Marsaglia, W.~W.~Tsang, and J.~Wang.
\newblock Evaluating Kolmogorov's distribution.
\newblock \textit{Journal of Statistical Software}, 8(18):1--4, 2003.
\newblock doi: https://doi.org/10.18637/jss.v008.i18

\bibitem{mclachlan2000}
G.~J.~McLachlan and D.~Peel.
\newblock \textit{Finite Mixture Models}.
\newblock John Wiley \& Sons, 2000.
\newblock doi: https://doi.org/10.1002/0471721182

\bibitem{burnham2002}
K.~P.~Burnham and D.~R.~Anderson.
\newblock \textit{Model Selection and Multimodel Inference: A Practical Information-Theoretic Approach}.
\newblock Springer, 2nd edition, 2002.
\newblock doi: https://doi.org/10.1007/b97636

\bibitem{kass1995}
R.~E.~Kass and A.~E.~Raftery.
\newblock Bayes factors.
\newblock \textit{Journal of the American Statistical Association}, 90(430):773--795, 1995.
\newblock doi: https://doi.org/10.1080/01621459.1995.10476572

\bibitem{virtanen2020}
P.~Virtanen et~al.
\newblock SciPy 1.0: Fundamental algorithms for scientific computing in Python.
\newblock \textit{Nature Methods}, 17(3):261--272, 2020.
\newblock doi: https://doi.org/10.1038/s41592-019-0686-2

\bibitem{hutto2014}
C.~Hutto and E.~Gilbert.
\newblock VADER: A parsimonious rule-based model for sentiment analysis of social media text.
\newblock In \textit{Proceedings of the International AAAI Conference on Web and Social Media (ICWSM)}, 8(1):216--225, 2014.
\newblock doi: https://doi.org/10.1609/icwsm.v8i1.14550

\bibitem{kleinbaum2012}
D.~G.~Kleinbaum and M.~Klein.
\newblock \textit{Survival Analysis: A Self-Learning Text}.
\newblock Springer, 3rd edition, 2012.
\newblock doi: https://doi.org/10.1007/978-1-4419-6646-9

\bibitem{burr1942}
I.~W.~Burr.
\newblock Cumulative frequency functions.
\newblock \textit{Annals of Mathematical Statistics}, 13(2):215--232, 1942.
\newblock doi: https://doi.org/10.1214/aoms/1177731607

\bibitem{tadikamalla1980}
P.~R.~Tadikamalla.
\newblock A look at the Burr and related distributions.
\newblock \textit{International Statistical Review}, 48(3):337--344, 1980.
\newblock doi: https://doi.org/10.2307/1402945

\bibitem{chen2013}
Z.~Chen and J.~Berger.
\newblock When, why, and how controversy causes conversation.
\newblock \textit{Journal of Consumer Research}, 40(3):580--593, 2013.
\newblock doi: https://doi.org/10.1086/671345

\bibitem{kleiber2003}
C.~Kleiber and S.~Kotz.
\newblock \textit{Statistical Size Distributions in Economics and Actuarial Sciences}.
\newblock John Wiley \& Sons, 2003.
\newblock doi: https://doi.org/10.1002/0471457175

\bibitem{hawkes1971}
A.~G.~Hawkes.
\newblock Spectra of some self-exciting and mutually exciting point processes.
\newblock \textit{Biometrika}, 58(1):83--90, 1971.
\newblock doi: https://doi.org/10.1093/biomet/58.1.83

\bibitem{iannone2019}
F.~Iannone et~al.
\newblock CRESCO ENEA HPC clusters: A working example of a multifabric GPFS Spectrum Scale layout.
\newblock In \textit{2019 International Conference on High Performance Computing \& Simulation (HPCS)}, pages 1051--1052, 2019.
\newblock doi: https://doi.org/10.1109/HPCS48598.2019.9188135

\end{thebibliography}

\end{document}